\documentclass[twocolumn,prl,a4paper,showpacs,floatfix,preprintnumbers,amsmath,amssymb]{revtex4}

\usepackage{graphicx}
\usepackage{dcolumn}
\usepackage{bm}

\newcommand{\be}{\begin{equation}}
\newcommand{\ee}{\end{equation}}
\newcommand{\wz}{\omega_{z}}
\newcommand{\wy}{\omega_{y}}

\newcommand{\w}{\omega}

\begin{document}


\title{Double-Slit Interferometry with a Bose-Einstein Condensate}

\author{L.A. Collins$^1$, L. Pezz\'e$^{1,2}$, 
A. Smerzi$^{1,2}$, G.P. Berman$^1$ and A.R. Bishop$^1$ }
\affiliation{1) Theoretical Division, Los Alamos National Laboratory,
Los Alamos, New Mexico 87545, USA\\
2) Istituto Nazionale per la Fisica della Materia BEC-CRS\\
and Dipartimento di Fisica, Universit\`a di Trento, I-38050 Povo, Italy}

\date{\today}

\begin{abstract}
A Bose-Einstein ``double-slit" interferometer has been recently
realized experimentally by
{\it Y. Shin {\em et. al.}, Phys. Rev. Lett. {\bf{92}} 50405 (2004)}. 
We analyze the interferometric steps by solving numerically
the time-dependent Gross-Pitaevski equation in three-dimensional space. We 
focus on the adiabaticity time scales of the problem and on 
the creation of spurious collective excitations  
as a possible source of the strong dephasing observed experimentally. 
The role of quantum fluctuations is discussed.
\end{abstract}

\maketitle

{\bf Introduction}.
Several current efforts in the field of dilute Bose-Einstein condensates (BEC)
are focusing on the creation of new technological devices, including
quantum computers \cite{rolston02} and ultrasensitive interferometers
\cite{kasevich02} to detect and measure weak forces.  Atom wave
interferometry already provides unprecedented sensitivities to detect
rotations, accelerations, and gravity gradients
\cite{berman_book}.  Performances can be further improved with
interferometers based on BEC, which are the highest brilliant coherent
sources of matter waves and which allow for larger
separations between different interferometric paths.

Interference between two spatially separated condensates was first demonstrated 
in \cite{andrews97}, and theoretically analyzed in \cite{rohrl97}. 
A BEC trapped in a harmonic magnetic trap 
was split in two symmetric halves by a laser knife. After releasing the
external fields, the interference pattern of the overlapping condensates
was observed with destructive imaging.
The measured relative phase, however, was not reproducible from shot to shot
but was randomly distributed due to the 
presence of a large noise in the
relative positions of the laser and the magnetic 
trap and of parasite currents when switching off the
magnetic fields. 
Reproducible interference patterns were eventually observed by trapping a
condensate in deep optical periodic potentials 
\cite{anderson98}. Neighboring wells of optical lattices, however, are
separated only by a fraction of a micron and cannot be individually addressed,
limiting their applications in technological devices.

\begin{figure}[t]
\begin{center}
\includegraphics[width=3.5in, height=1.7in]{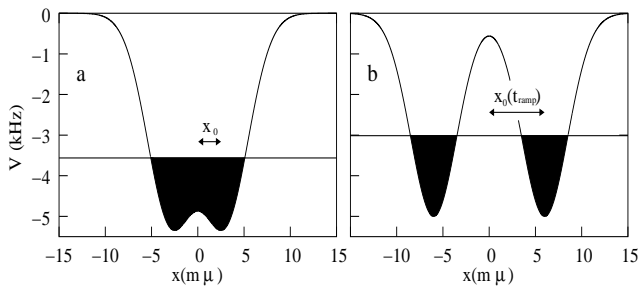}
\caption{Double Well potential in the $x$ direction. 
a) The initial condensate with a
distance between the two wells of $x_0$ = 3 $\mu m$ 
and the chemical potential higher than the barrier height.  
b) Displaced wells
in a time $t=t_{ramp}$ with $x_0(t_{ramp}) =6$ $\mu m$. In this 
configuration we have two independent condensates.}\label{Figdw}
\end{center}
\end{figure}

Recently, a stable double-well trap has been created in \cite{shin04}.
A single collimated laser beam was split with
an acoustic-optical modulator, and finally focused by a lens. 
A single, cigar-shaped condensate was trapped in a double-well 
having a barrier much smaller than the BEC chemical potential, 
see Fig. (\ref{Figdw}a).
The condensate was then split along the axial direction 
by linearly increasing, in a ramping time
$t_{ramp}$, both the distance between the two
wells and the height of the interwell barrier, see Fig. (\ref{Figdw}b). 
The final distance between the two condensates was
$\sim 12~\mu$m, allowing for $individual$
addressing and manipulation. 
After holding the two condensates in the respective 
traps for a time 
$t_{hold}$, the confining field was turned off. 
The interference pattern of the two overlapping condensates 
was measured by destructive techniques and
was reproducible in different realizations of the experiment. 
However,
a loss of coherence was observed when the condensates 
were held in the separated wells longer than $t_{hold} \approx 5$ ms.        
This strong dephasing has been tentatively attributed \cite{shin04} 
to axial and breathing mode excitations
created during the splitting of the condensate.
Attempts to increase the adiabaticity of the process with larger
separation times $t_{ramp}$ did not 
improve the stability of the measured phase. 
The dephasing manifested itself as a gradual decrease of the 
phase contrast accompanied by
a bending and kinking of the interference pattern, which eventually makes
the phase measurement impossible.

In this Letter we theoretically analyze the MIT experiment 
as a prototype of a general BEC interferometer. Our analisys  
is quite general, and is relevant, for instance, 
in the study of the splitting and recombination 
of BEC propagating in atom-chip wave guides \cite{atomchip}.
We focus on the role of the collective excitations created during the
splitting process as a possible dephasing mechanism.
We numerically solve the full three-dimensional
time-dependent Gross-Pitaevskii (GP) equation to reproduce the
interferometric steps realized in the experiment. 
We study the adiabaticity of the splitting process
and predict the interference contrast as a function of 
the various time scales of the problem. 
We finally consider the corrections to the GP dynamics
arising from quantum fluctuations.

{\bf Protocol.}
Our simulation of the interferometric experiment consists of four steps:
i) {\it Initialization.} We load a BEC into an optical double well potential 
of gaussian shape in the $x$ direction and of 
harmonic shape in the $y$ and $z$ directions 
with frequencies $\wy$ and $\wz$, respectively.  
We fix an initial separation $x_0$ between the wells 
in such a way that the height of the potential barrier is much smaller than
the chemical potential, Fig. (\ref{Figdw}a). 
We solve the GP equation in imaginary time in order to
find the ground state (GS) of the system.
ii) {\it Separation.} We separate the wells by ramping 
linearly from $x_0$ at time $t=0$ to
$x_{ramp}$ at $t _{ramp}$, Fig. (\ref{Figdw}b). 
iii) {\it Holding.} Once the wells are separated, we generally allow the
wavefunction to evolve in the trap for a time $t_{hold}$.
iv) {\it Ballistic (free) propagation.} After a time
$t =t_{ramp} + t_{hold}$, we simply release the trapping potential
and allow the packets from 
the separated wells to merge and overlap, generating an
interference pattern. The condensates are finally 
imaged after a time $t_{free}$ 
of ballistic expansion by integrating along the 
$y$ direction, simulating in this way
the data collected experimentally.      
We cast the interference problem 
into a three-dimensional Gross-Pitaevskii (GP) equation.
We introduce scaled units with the energy in $\hbar\omega_z$,
length in 
$d_z = \sqrt{\frac{\hbar}{m\omega_z}}$, and time in ${\omega_z}^{-1}$, 
obtaining 
\begin{equation}  \label{GP}
i\frac{\partial}{\partial{t}} \psi({\bm{r}},t) = \Big[-\frac{1}{2}{\nabla}^2 + V_{ext}({\bm{r}},t) + V_{NL}({\bm{r}},t)\Big]
\psi({\bm{r}},t),
\end{equation}
where $V_{NL}({\bm{r}},t) =
4\pi N \frac{a}{d_z}|\psi({\bm{r}},t)|^2$ is the nonlinear 
potential arising from the interatomic interaction, 
with $a$ the scattering length and $N$ the number of
condensate atoms in the trap.
The external potential $V_{ext}({\bm{r}},t)$ 
is given by a combination of a harmonic confinement
along the $y,z$-directions, 
$V_{ho}(y,z) = \frac{1}{2} \bigg(\frac{\wy^2}{\wz^2}y^2 + z^2\bigg)$,
and a double-well
time-dependent gaussian confinement along the $x$ direction
$V_{dw}(x,t) =  -\frac{V_0}{\hbar \wz}\Big[ e^{-(x - x_0(t))^2/2{\sigma}^2} + 
e^{-(x + x_0(t))^2/2{\sigma}^2}\Big]$, 
where $2x_0(t)$ is the distance of the potential 
wells at time t, and $\sigma$ is 
the width of each gaussian well.
We base our simulations on the MIT experiment \cite{shin04} 
considering $N \sim 5 10^5$ $^{23}$Na atoms. The interatomic 
scattering length is $a = 2.8$ nm, and the trap parameters are 
$\wy=615 \times 2\pi$ Hz, $\wz=30 \times 2\pi$ Hz (giving a length unit
$d_z$ = 3.8 $\mu$m), 
$V_0$  =  $h \times 5 $  kHz, 
and $\sigma  =  2.5$ $\mu$m. 
The bottom portion of each gaussian well approximates a
harmonic potential with frequency 
$\w^{h}_{x} =\sqrt{\frac{V_0}{m{\sigma}^{2}}} = 2 \pi \times 593$ Hz
In  the initial configuration, Fig.(\ref{Figdw},a),  
$x_0=3$ $\mu m$ and the chemical potential $\mu=h \times 1.78$ kHz. 
The wells are then separated to a distance $x_0 = 6~\mu$m, 
Fig. (\ref{Figdw},b), creating two condensates each with 
chemical potential $ \mu=h \times 2.04$ kHz.

{\bf Results and Discussion}.
We performed GP simulations  
by varying the parameters $t_{ramp}$ and $t_{hold}$
keeping fixed the expansion time $t_{free} = 5.3$ ms \cite{comment2}.
The central goal of this section is to extract the relevant adiabaticity time
scale of the double-slit BEC interferometer.
This requires studying the excitations created during 
the splitting of the trap and how these propagate during the
holding time.

\begin{figure}
\begin{center}
\includegraphics[width=3.5in]{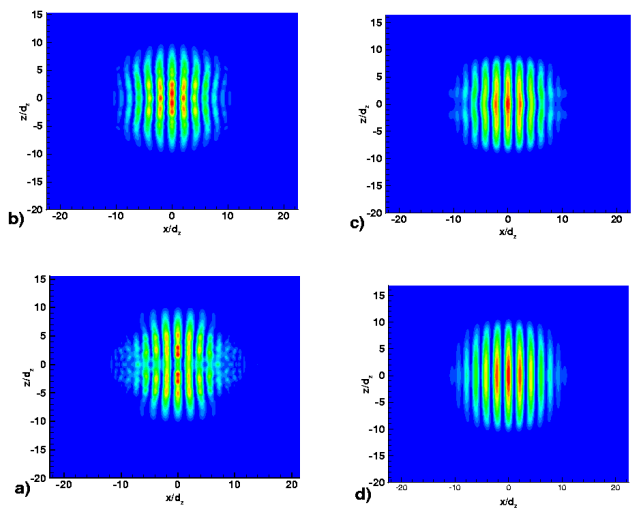}
\vspace{.2cm}\caption{Contour plots of the probability density
integrated over the $y$-coordinate ${P_{y}}(x,z,t)$ from 3D simulations for $t_{hold}$= 10.6 ms 
and $t_{free}$ = 5.3 ms, as a function of the ramping time $t_{ramp}$: a) 2.7 ms, b) 5.3 ms, 
c) 10.6 ms, and d) 21.2 ms.}
\label{Fig-ramptw2c}
\end{center}
\end{figure}

{\it Ramp time}.
We first examine the behavior of the system as a function of the 
separation time, $t_{ramp}$, for fixed $t_{hold}$ = 10.6 ms. 
In Fig. (\ref{Fig-ramptw2c}), we show
the contours of the $y$-integrated probability density $P_{y}(x,z,t)$,
which closely resemble the projected images of the experiment.
For short separation times we have observed 
considerable distortions and dephasing in the
interference patterns.

Given the symmetries of the system, 
low-lying excitations
can only have even parity. For example, in the axial direction
the longest oscillation period is
$\tau_z \sim {{\pi}\over {\omega_z} } = 16.6$ ms, and faster 
ramping times can easily excite large amplitude monopole and 
quadrupole oscillations.
However, we have numerically checked that such collective modes 
retain a high degree of collectivity (due to the harmonicity of the
trap) and  
cannot be a source of dephasing. 
Nonlinear couplings, which can potentially destroy the interference patterns, 
apparently occur at much longer times than those considered
in the experiment.

We now consider 
the frequencies of low-lying even excitations along the radial direction
as a function of the interwell distances.
The oscillation periods first increase along with the increase 
of the interwell barrier height. These reach a maximum value 
when the difference between the energy of the
interwell barrier and the chemical potential
becomes equal to the frequency of the first odd collective mode,
corresponding to the Josephson ``plasma" oscillation. 
Further increasing the interwell distance results in the periods
decreasing to a final saturation 
once the two wells are completely separated.
This behaviour reflects the evolution of the
single particle energies of the double well during the ramping process.  
The maximum oscillation period sets the $relevant$ adiabaticity time scale 
of the double-slit BEC interferometer.

In each separate well, both even and odd excitations modes
can be excited along the radial axis.
The lowest in energy is the dipole collective mode, whose oscillation
frequency can be calculated in a variational approach to be
$\omega_D^2 = {\omega^h_x}^2 ({1 \over {1 + \gamma^2 / \sigma^2}})^{(3/2)}$,
with $\gamma$ the the radial width of the
condensate.
The trapping potential
is highly anharmonic and the correponding 
oscillations frequencies strongly depend on the
chemical potential of the system, even in the Thomas-Fermi limit 
(contrary to what happens with an harmonic
confinement). 
With $N = 10^5$, the largest frequency is obtained at an 
interwell distances $x_0 \sim 3.7~\mu$m and chemical potential
$\mu \simeq 0.64~V_{dw}(x=0)$: 
$\omega_{min} = 0.65~\omega_x^h$,
and the oscillation period
is $\tau_{max} = 1.54~\tau_x^h$.
For comparison, when
$x_0=6~\mu$m, and the two wells are completely separated,
$\omega_D = 0.75~\omega_x^h$
and the oscillation period
is $\tau_D = 1.33~\tau_x^h = 2.58$ms.

The breaking down of adiabaticity, however,
is not the only source of dephasing.
We have not observed any loss of visibility
after imposing harmonic
confinement along the $x$ axis.
On the other hand, we have found that, in the presence of the
gaussian confinement,
the interference pattern quickly degrades while increasing the
number of atoms, keeping all other parameters fixed.
Fast rampings induce
large amplitude oscillations along the radial direction of each trap.
Due to the combined anharmonicity of the confining potential
and the non-linearity
arising from the interatomic interaction, the energy of the
excitation
quickly redistributes among different modes, rapidly
damping the collective oscillation.
We have not observed any coupling of such modes with oscillations among
the radial axis. It is useful to recall that
the total energy of the system remains conserved in the GP
simulation, and the effective damping is only due to redistribution
of the collective energy
among different Bogoliubov excitations levels.
With smaller oscillation amplitudes, induced with longer ramping times,
the visibility of the fringes
has been greatly improved Fig.(\ref{Fig-ramptw2c}).
However, this effect is in contrast with the experimental
results \cite{shin04} where an
increasing of the ramping time, even well beyond $\tau_{max}$,
does not improve the interference pattern,
which degradates for $t_{hold}\ge 5$ ms (see Fig. 3 in \cite{shin04}),
independently of $t_{ramp}$.

\begin{figure}
\begin{center}
\includegraphics[width=3.5in]{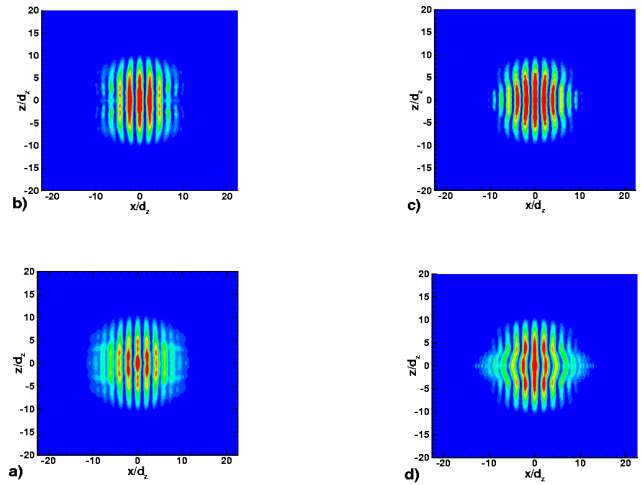}
\vspace{.2cm}\caption{Contour plots of the probability density
integrated over the $y$-coordinate ${P_{y}}(x,z,t)$  for $t_{ramp}$= 5.3 ms 
and $t_{free}$ = 5.3 ms as a function of the hold time $t_{hold}$: 
a) 0, b) 5.3 ms, c) 10.6 ms, 
and d) 21.2 ms.}
\label{Fig-ramp1tw}
\end{center}
\end{figure}

{\it Hold Time}.
We now investigate the role of the hold times $t_{hold}$
in the deterioration of the interference pattern. 
Fig.(\ref{Fig-ramp1tw}) presents contours for
the $y$-integrated probability density for different $t_{hold}$
and fixed $t_{ramp} = 5.3$ ms. 
We clearly note a growing deformity of the patterns as
the confinement time increases, along with the redistribution of 
the initial energy among various modes. The maximum dephasing indeed occurs
when the oscillation is completely damped. 
Our simulation goes in the direction of the experiment:
an increasing of $t_{hold}$ corresponds to a growing 
distortion of the fringes pattern. 
However, there is an important difference. 
We do not observe  
a complete distortion of the fringes pattern for $t_{hold} \ge 5$ ms,
as in \cite{shin04}.
Instead, we find distorted but still defined fringes even 
for $t_{hold} \approx 20$ ms.
Moreover, such distortion 
can eventually be reduced by increasing the ramping time,
as suggested by Fig.(\ref{Fig-ramptw2c}). 
Changing the relative depth of the two wells 
to include an asymmetry in the x-direction, has not lead to 
a more rapid deterioration of the fringe contrast.

Another important difference between the experimental findings and our
GP simulations occurs in the shape 
of the distorted interference pattern. We  
always find symmetric bending or kinking of the fringes, 
contrary to that observed in Fig. (3) of \cite{shin04}. 
The asymmetric kinking of the experiment suggest the presence of
slight additional geometry variations of the two traps.
In the real experiment, indeed, the shape as well as
the parameters of the trap can be slightly different from the corresponding
mathematical form we have used in our simulations.
A small disalignement of the two traps, for instance, can induce
small amplitude out of phase dipole oscilations along the $z$
axis which can originate an asymmetric bending. 
Residual excitations due to the transportation of the condensate
to the ``science chamber" \cite{shin04} prior the initialization step i),
might also be the source of the dephasing observed experimentally.
\cite{aaron}.
We can conclude this part by stating that the excitations
coming from the creation of 
two separated condensates give
significant distortion and loss of contrast in the interference fringes, 
but they do not destroy completely the pattern. 
In general such distortions increase with $t_{hold}$, as in the experiment,
and with the chemical potential, but 
it can be substantially reduced by decreasing the ramping time $t_{ramp}$. 

{\bf Beyond Gross-Pitaevskii.}
The Gross-Pitaevskii framework
does not account for dephasing arising from quantum (many-body)
correlations. While, on the one hand, a full calculation of the quantum
dynamics is well beyond current computational capabilities,  
simple estimates 
\cite{imamoglu97,javanainen99,smerzi00,menotti01}
of the dynamical evolution of the fringe contrast
can be carried out within the two-mode approximation and 
tested experimentally. The quantum phase 
Hamiltonian governing the condensate is (setting $\hbar =1$) \cite{anglin01}:
\begin{equation}
i \frac{\partial \Psi }{\partial t}=
\left[ -\frac{E_{c}}{2}\frac{\partial ^{2}}{\partial \phi ^{2}}
-E_{j} \cos {\phi }-\frac{E_{j}^{2}}{N^{2}E_{c}}\cos {2\phi }\right] \Psi, 
\label{qpm}
\end{equation}
where $\Psi(\phi,t)$ is the probability amplitude for the
relative phase $\phi$ of the two condensates. The ``charging" energy
$E_c = 2 \frac{\partial \mu }{\partial N}$ remains approximately constant
during the splitting of the condensate, while
the Josephson coupling energy $E_j$ decreases exponentially  
with the interwell distance.
In the strong coupling limit ${E_j / E_c} >> 1$ (achieved when
the chemical potential is close enough to the interwell barrier),
the phase oscillation frequency is
$\omega_j = \sqrt{E_c E_j + 4 E_j^2 / N^2 }$. 
The phase probability $|\Psi(\phi,t)|^2$ has a gaussian distribution with
dispersion $\sigma^2 = {1 \over 2} {E_c \over \omega_j} << 1$.
As long as $\sigma(t)$
remains small during the dynamics,
the system can be described in the GP framework.
With a linear ramping, $d(t) = 2 x_0 t / \tau_{ramp}$,
and in the $WKB$ approximation, the Josephson coupling energy is
$E_j(t) \sim e^{(- t / \tau_{eff})}$, 
with $\tau_{eff} = \tau_{ramp} / S$ and effective 
action $S=2 x_0 \sqrt{2 m (V_0 -\mu)}$.
At a first stage, while $E_j$ decreses with time, the amplitude $\Psi(\phi,t)$
follows adiabatically the ground state of the effective quantum potential
in Eq.(\ref{qpm}). Breakdown of adiabaticity occurs at the freeze-out time
$t_f$ given by
$\omega_j(t_f) \simeq {{2 \pi} \over \tau_{eff}}$, namely when the 
Josephson period becomes equal to the characteristic time of the change of the
Josephson coupling energy 
$\tau_{eff} = - (\frac{d E_j }{d t} / E_j)^{-1}$.
At longer times the dynamics is dominated by
the kinetic part of the quantum Hamiltonian Eq.(\ref{qpm}). 
The temporal evolution 
of the phase fluctuations is therefore given by 
$\sigma^2(t) = \sigma^2(t_f) + E_c^2/4 \sigma^2(t_f) t^2$, with $\sigma(t_f)$
being the width calculated at the freeze-out time. Replacing the corresponding
expressions, we finally obtain:
\begin{equation}
\sigma^2(t) = {E_c \over 2} (\omega_j(t_f)^{-1} + \omega_j(t_f) t^2 )
\simeq {E_c \over 2} 
( {\tau_{ramp}\over {2 \pi S}} + {2 \pi S}  { t_{hold}^2 \over \tau_{ramp}}), 
\label{phasedyn}
\end{equation}
which is the central result of this section.
To experimentally test the quantum phase dynamics Eq.(\ref{phasedyn}) 
it would be necessary to average over 
several identical interferometric realizations. In each experiment, 
an interference pattern
will actually be observed \cite{castin97}, but with a relative phase
chosen randomly with
a gaussian distribution of width given by Eq.(\ref{phasedyn}). 
The quantum dynamics will therefore manifest itself 
as a loss of fringe visibility 
of the ensemble averaged interference patterns, which will be 
almost completely washed out
when $\sigma \sim 2$.
Replacing the experimental values of \cite{shin04}, 
this happens with a holding time 
$t_{hold} \simeq \sqrt{ {{4 \hbar^2}\over {\pi S E_c}} 
\tau_{ramp}} \simeq 2 \pi 50$ ms. 

{\bf Conclusions.}
We have theoretically studied the double-slit interferometer recently
created in \cite{shin04} solving numerically
the full 3D time-dependent Gross-Pitaevski equation.
We have studied the adiabaticity time scales of the system, 
concluding that the dephasing arising from the creation of 
spurious excitations can be strongly reduced 
by increasing the ramping time of the double-well.
Such findings indicate that the
loss of visibility and the bending of the interference pattern
observed experimentally should arise from a different noise source.

{\bf Acknowledgements.}
We thank A.E. Leanhardt for very useful comments and for sharing 
results of his unpublished Ph.D. thesis. 
This work has been partially supported under the auspices of the U.S
Department of Energy through the Theoretical Division at the Los
Alamos National Laboratory.


\begin{thebibliography}{99}
\bibitem{comment2}
The fixing of the grid limits our simulation to a free expansion time 
$t_{free} \approx 5$ msec. At such an expansion time the two clouds are 
completely overlapped, and we do not expect any qualitative difference with
longer expansion times. Complementary 2D GP simulations and
semi-analytical Thomas-Fermi approaches extended up to $t_{free} = 30$ ms
confirm the full 3D GP trends.
\end{thebibliography}
\end{document}